# Bioinformatics and biomedical informatics with ChatGPT: Year one review┼

Jinge Wang[1], Zien Cheng[1], Qiuming Yao[2], Li Liu[3,4], Dong Xu[5], Gangqing Hu[1,*]
[1]Department of Microbiology, Immunology & Cell Biology, West Virginia University, Morgantown, WV 26506, USA
[2]School of Computing, University of Nebraska-Lincoln, Lincoln, NE 68588, USA
[3]College of Health Solutions, Arizona State University, Phoenix, AZ 85004, USA
[4]Biodesign Institute, Arizona State University, Tempe, AZ 85281, USA
[5]Department of Electrical Engineer and Computer Science, Christopher S. Bond Life Sciences Center, University of Missouri, Columbia, MO 65211, USA

Running title: Bioinformatics with ChatGPT

*Corresponding author: Gangqing Hu: Michael.hu@hsc.wvu.edu

**ABSTRACT**

The year 2023 marked a significant surge in the exploration of applying large language model (LLM) chatbots, notably ChatGPT, across various disciplines. We surveyed the applications of ChatGPT in bioinformatics and biomedical informatics throughout the year, covering omics, genetics, biomedical text mining, drug discovery, biomedical image understanding, bioinformatics programming, and bioinformatics education. Our survey delineates the current strengths and limitations of this chatbot in bioinformatics and offers insights into potential avenues for future developments.

## 1. INTRODUCTION

In recent years, artificial intelligence (AI) has attracted tremendous interest across various disciplines, emerging as an innovative approach to tackling scientific challenges [1]. The surge in data generated from both public and private sectors, combined with the rapid advancement in AI technologies, has facilitated the development of innovative AI-based solutions and accelerated scientific discoveries [1, [2, [3]. The launch of the Chat Generative Pre-trained Transformer (ChatGPT) to the public towards the end of 2022 marked a new era in AI. The biomedical research community embraces this new tool with immense enthusiasm. In 2023 alone, at least 2,074 manuscripts were indexed in PubMed when searching with the keyword "ChatGPT". These studies demonstrate that ChatGPT and similar models have great potential to transform many aspects of education, biomedical research, and clinical practices [4, [5, [6, [7].

The core of ChatGPT is a large-language model (LLM) trained on a vast corpus of text and image materials from the internet, including biomedical literature and code [8]. Its ability to comprehend and respond in natural language positions ChatGPT as a valuable tool for biomedical text-based inquiry [9]. Particularly noteworthy is its potential in assisting bioinformatics analysis, enabling scientists to conduct data analyses via verbal instructions [10, [11, [12]. Surprisingly, a search on PubMed using the keywords "ChatGPT" and "bioinformatics" returned only 30 publications. While this number could have been underestimated by limiting the search to PubMed, and a few hundred related articles are probably archived as preprints or under review, it still suggests that the application of this innovative tool in bioinformatics is relatively underexplored compared to other areas of biomedical research.

In this review, we summarize recent advancements, predominantly within 2023, in the application of ChatGPT across a broad spectrum of bioinformatics and biomedical informatics topics, including omics, genetics, biomedical text mining, drug discovery, biomedical images, bioinformatics programming, and bioinformatics education (**Figure 1**). As the topics are relatively new, this survey included not only publications in journals but also preprints in various archive platforms. Our objective is to encapsulate recurring themes from independent works within the same topic or across multiple topics, pinpointing prospective avenues for further exploration. Additionally, this review allows us to identify challenges in integrating chatbots into bioinformatics, such as inefficiency in prompt generation, uncertainty in responses, and concerns over data privacy [13, [14, [15]. The insights from this analysis are anticipated to benefit other domains where the integration of chatbot technology is actively pursued.

## 2. LITERATURE SELECTION

We searched Google Scholar, PubMed, and various preprint servers (aRxiv, bioRxiv, medRxiv, chemRxiv, and Research Square) using keywords such as "ChatGPT" in combination with "Bioinformatics", "Computational Biology", "Genetics", "Text mining", or "Drug Discovery." We then reviewed titles and abstracts to select papers that use ChatGPT in bioinformatics and biomedical informatics. We also utilized backward and forward citation tracking for each identified publication to expand the pool, resulting in 65 manuscripts. Lastly, we excluded poster manuscripts and manuscripts lacking in-depth analysis. In the end, we narrowed down to 62 research articles (**Supplementary Table S1**).

Most of the works reviewed (72.6%) are about performance assessment, while 19.3% lean towards direct applications of GPT (**Supplementary Table S1**). A significant proportion of the works (45 out of 62) were initially released as preprints, reflecting the emergent nature of this field. In addition to highlighting findings from individual works, we also identified findings supported across multiple independent studies. Out of the 22 preprints that were not formally published at the time of writing, 17 preprints contribute to this direction (**Supplementary Table S1**). This cross-validation among studies strengthens the reliability of observed trends and shared insights, especially for findings described in preprints.

## 3. OMICS

Omics techniques are extensively employed in biomedical research, generating vast amounts of data that necessitate careful analysis to uncover significant discoveries. A novel application of GPT-4 is to annotate cell types in single-cell RNA sequencing data [16], traditionally a labor-intensive and expertise-demanding task. Leveraging the wealth of online texts that offer detailed descriptions of signature genes for various cell types, GPT-4 can efficiently identify cell types based on a tissue name and a list of marker genes, as few as ten, identified from standard single-cell analysis pipelines such as Seurat **(Figure 2)**. When evaluated across ten datasets encompassing hundreds of tissues and cell types, GPT-4 demonstrates strong concordance with manual annotations and surpasses several competing methods including CellMarker 2.0, ScType, and SingleR [16]. GPT-4 achieves this impressive performance with basic prompts and does not require biology expert, referencing data sets, or coding experience, thus making cell type annotation easily accessible to general biomedical researchers for scRNA-Seq data analysis. However, given the undisclosed nature of GPT's training data and the potential for AI-generated errors, expert validation is recommended before leveraging its annotations in further research, especially for tissues and cell types that are not widely studied.

Evaluating GPT models in genomics necessitates benchmark datasets with established ground truths. GeneTuring [17] serves this role with 600 questions related to gene nomenclature, genomic locations, functional characterization, sequence alignment, etc. When tested on this dataset, GPT-3 excels in extracting gene names and identifying protein-coding genes, while ChatGPT (GPT-3.5) and New Bing show marked improvements. Nevertheless, all models face challenges with SNP and alignment questions [17]. This limitation is effectively addressed by GeneGPT [18], which utilizes Codex to consult the National Center for Biotechnology Information (NCBI) database.

## 4. GENETICS

In North America, 34% of genetic counselors incorporate ChatGPT into their practice, especially in administrative tasks [19]. This integration marks a significant shift towards leveraging AI for genetic counseling and underscores the importance of evaluating its reliability. Doung and Solomon [20] analyzed ChatGPT's performance on multiple-choice questions in human genetics sourced from Twitter. The chatbot achieves a 70% accuracy rate, comparable to human respondents, and excels in tasks requiring memorization over critical thinking. Further analysis by Alkuraya, I. F. [21] revealed ChatGPT's limitations in calculating recurrence risks for genetic diseases. A notable instance involving cystic fibrosis testing showcases the chatbot's ability to derive correct equations but falter in computation, raising concerns over its potential to mislead even professionals [21]. This aspect of plausible responses is also identified as a significant risk by genetic counselors [19].

These observations have profound implications for the future education of geneticists. It indicates a shift from memorization tasks to a curriculum that emphasizes critical thinking in varied, patient-centered scenarios, scrutinizing AI-generated explanations rather than accepting them at face value [22]. Moreover, it stresses the importance of understanding AI tools' operational mechanisms, limitations, and ethical considerations essential in genetics [20]. This shift prepares geneticists better for AI use, ensuring they remain informed on the benefits and risks of technology.

## 5. BIOMEDICAL TEXT MINING

For biomedical text mining with ChatGPT, we first summarized works that evaluate the performance of ChatGPT in various biomedical text mining tasks and compared it to state-of-the-art (SOTA) models. Then, we explored how ChatGPT has been used to reconstruct biological pathways and prompting strategies used to improve the performance.

### 5.1. PERFORMANCE ASSESSMENTS ACROSS TYPICAL TASKS

Biomedical text mining tasks typically include name entity recognition, relation extraction, sentence similarity, document classification, and question answering. Chen, Q., *et al.* [23] assessed ChatGPT-3.5 across 13 publicly available benchmarks. While its performance in question answering closely matched SOTA models like PubmedBERT [24], ChatGPT-3.5 showed limitations in other tasks, with similar observations made for ChatGPT-4 [7, [25, [26]. Extensions to sentence classification and reasoning revealed that ChatGPT was inferior to SOTA pretrained models like BioBERT [27]. These studies highlight the limitations of ChatGPT in some specific domains of biomedical text mining where domain-optimized language models excel. Nevertheless, when the training sets with task-specific annotations are not sufficient, zero-shot LLMs, including ChatGPT-3.5 outperform SOTA finetuned biomedical models [28]. A compilation of performance metrics for ChatGPT and other baseline models on various biomedical text mining tasks is listed in **Supplementary Table S2**.

Biomedical Knowledge Graphs (BKGs) have emerged as a novel paradigm for managing large-scale, heterogeneous biomedical knowledge from expert-curated sources. Hou, Y*., et al.* [29] evaluated ChatGPT's capability on question and answering tasks using topics collected from the "Alternative Medicine" subcategory on "Yahoo! Answers" and compare to the Integrated Dietary Supplements Knowledge Base (iDISK) [30]. While ChatGPT-3.5 showed comparable performance to iDISK, ChatGPT-4 was superior to both. However, when tasked to predict drug or dietary supplement repositioned for Alzheimer's Disease, ChatGPT primarily responded with candidates already in clinical trials or existing literature. Moreover, ChatGPT's efforts to establish associations between Alzheimer's Disease and hypothetical substances were less than impressive. This highlights ChatGPT's limitations in performing novel discoveries or establishing new entity relationships within BKGs.

ChatGPT's underperformance in some specific text mining tasks against SOTA models or BKGs identifies areas for enhancement; On the other hand, finetuning LLMs, although beneficial, remains out of reach for most users due to the high computational demand. Therefore, techniques like prompt engineering, including one/few-shot in-context learning and Chain-of-Though (CoT; See **Table 1** for terminologies cited in this review), can be more practical to improve LLM efficiency in text mining tasks [23, [25, [27, [31]. For instance, incorporating examples with CoT reasoning enhances the performances of ChatGPT over both zero-shot (no example) and plain examples in sentence classification and reasoning tasks [27] as well as knowledge graph reconstruction from literature titles [32]. However, simply increasing the number of examples does not always correlate with better performance [25, [27]. This underscores another challenge in optimizing LLMs for specialized text mining tasks, necessitating more efficient prompting strategies to ensure consistent reliability and stability.

## 5.2. BIOLOGICAL PATHWAY MINING

Another emerging application of biomedical text mining from LLMs is to build biological pathways. Azam, M*., et al.* [33] conducted a broader assessment of mining gene interactions and biological pathways across 21 LLMs, including seven Application Programming Interface (API)-based and 14 open-source models. ChatGPT-4 and Claude-Pro emerged as leaders, though they only achieved F1 scores less than 50% for gene relation predictions and a Jaccard index less than 0.3 for pathway predictions. Another evaluation work on retrieving protein-protein interaction (PPI) from sentences reported a modest F1 score for both GPT-3.5 and GPT-4 with base prompts [34]. All the studies underscore the inherent challenges generic LLMs face in delineating gene relationships and constructing complex biological pathways from biomedical text without prior knowledge or specific training.

The capabilities of ChatGPT in knowledge extraction and summarization present promising avenues for pathway database curation support. Tiwari, K*., et al.* [35] explored its utility in the Reactome curation process, notably in identifying potential proteins for established pathways and generating comprehensive summaries. For the case study on the circadian clock pathway, ChatGPT proposed 13 new proteins, five of which were supported by the literature but overlooked in traditional manual curation. When summarizing pathway from multiple literature extracts, ChatGPT struggled to resolve contradictions, but gained improved performance when inputs contained in-text citations. Similarly, the use of ChatGPT for annotating long non-coding RNAs in the EVLncRNAs 3.0 database [36] faces issues with inaccurate citations. Both works emphasize cautions on direct use of ChatGPT in assisting in database curation.

Supplementing ChatGPT with domain knowledge or literature has been shown to mitigate some of its intrinsic limitations. The inclusion of a protein dictionary in prompts improves performance for GPT-3.5 and GPT-4 in PPI task [34]. Chen, X*., et al.* [37] augmented ChatGPT with literature abstracts to identify genes involved in arthrofibrosis pathogenesis. Similarly, Fo, K*., et al.* [38] supplied GPT-3.5 with plant biology abstracts to uncover over 400,000 functional relationships among genes and metabolites. This domain knowledge/literature-backed approach enhances the reliability of chatbots in text generation by reducing AI hallucination [39, [40].

Addressing LLMs' intrinsic limitations can also involve sophisticated prompt engineering. Chen, Y., *et al.* [41] introduced an iterative prompt optimization procedure to boost ChatGPT's accuracy in predicting gene-gene interactions, utilizing KEGG pathway database as a benchmark. Initial tests without prompt enhancements showed a performance decline along with ChatGPT's upgrades from March to July in 2023, but the strategic role and few-shot prompts significantly countered this trend. The iterative optimization process, which employed the tree-of-thought methodology [42], achieved notable improvements in precision and F1 scores [41]. These experiments demonstrate the value of strategic prompt engineering in aligning LLM outputs with complex biological knowledge for better performance.

## 6. DRUG DISCOVERY

Drug discovery is a complex and failure-prone process that demands significant time, effort, and financial investment. The emerging interest in ChatGPT's potential to facilitate drug discovery has captivated the pharmaceutical community [43, [44, [45, [46]. Recent studies have showcased the chatbot's proficiency in addressing tasks related to drug discovery; a compilation of performance metrics for ChatGPT and other baseline models is listed in **Supplementary Table S3**. GPT-3.5, for example, has been noted for its respectable accuracy in identifying associations between drugs and diseases [47]. Furthermore, GPT models exhibit strong performance in tasks related to textual chemistry, such as generating molecular captions, but face challenges in tasks that require accurate interpretation of the Simplified Molecular-Input Line-Entry System (SMILES) strings [48]. Research by Juhi, A., *et al.* [49] highlighted ChatGPT's partial success in predicting and elucidating drug-drug interactions (DDIs). When benchmarked against two clinical tools, GPT models achieved an accuracy rate of 50-60% in DDI prediction and improved furhter by 20-30% with internet search through BING; a comparison to SOTA methods was not conducted [50]. When evaluated using the DDI corpus [51], ChatGPT achieved an micro F1 score of 52%, lower than SOTA BERT-based models [23]. In more rigorous assessments, ChatGPT was unable to pass various pharmacist licensing examinations [52, [53, [54]. It also shows limitations in patient education and in recognizing adverse drug reactions [55]. These findings suggest that, although ChatGPT offers valuable support in drug discovery, its capacity to tackle complex challenges is ineffective and necessitates close human oversight.

In the following few sections, we will review three important aspects of using LLM-chatbots such as ChatGPT in drug discovery (**Figure 3**). We first focused examples and tools that facilitate a human-in-the-loop approach for reliable use of ChatGPT in drug discovery. Then we highlighted the advances brought by strategic prompting using in-context learning with examples to increase response accuracy of ChatGPT. Lastly, we summarize the progress of using task- and or instruction finetune to adapt a foundational model to specific tasks, though demonstrated mostly by open-source models but could be extended to GPT-3.5 and GPT-4.

### 6.1. HUMAN-IN-THE-LOOP

The application of AI in drug development necessitates substantial expertise from human specialists for result refinement. This collaborative approach is illustrated in a case study focusing on the development of anti-cocaine addiction drugs aided by ChatGPT [56]. Throughout this process, GPT-4 assumes three critical roles in sparking new ideas, clarifying methodologies, and providing coding assistance. To enhance its performance, the chatbot is equipped with various plugins at each phase to ensure deeper understanding of context, access to the latest information, improved coding capabilities, and more precise prompt generation. The responses generated by the chatbot are critically evaluated with existing literature and expert domain knowledge. Feedback derived from this evaluation is then provided to the chatbot for further improvement. This iterative, human-in-the-loop methodology led to the identification of 15 promising multi-target leads for anti-cocaine addiction [56]. This example underscores the synergistic potential of human expertise and AI in advancing drug discovery efforts.

Several tools leveraging LLMs offer interactive interfaces to enhance molecule description and optimization. ChatDrug [57] is a framework that can use GPT API or other open source LLMs to streamline the process of editing small molecules, piptides, or proteins (**Figure 4**). It features a prompt design module equipped with a collection of template prompts customized for different types of editing tasks. The core of ChatDrug is a retrieval and domain feedback module to ensure that the response is grounded in real-world examples and safeguarded through expert scrutiny: The retrievel sub-module selects examples from external databases, while the domain feedback sub-module integrates feedback from domain experts through iteration. Additionally, ChatDrug includes a conversational module dedicated to further interactive refinement. Similar tools though based on other LLMs have been developped. DrugChat based on Vicuna-13b [58] offers interactive question-and-answer and textual explanations starting from drug graph representations. DrugAssist [59] based on Llama2-7B utilizes external database retrieval for hints and allowing iterative refinement with expert feedback. This process of iterative refinement, supported by example retrieval from external databases as contextual hints, also known as retrieval-augmented generation (RAG), and expert feedback enhances the model's accuracy and relevance to practical applications.

## 6.2. IN-CONTEXT LEARNING

In-context learning (ICL) enhances chatbots' responses by leveraging examples from a domain knowledgebase through prompting without finetuning a foundation model [60]. This approach utilizes examples closely aligned with the subject matter to ground the responses of ChatGPT with relevant domain knowledge [57, [61]. Evaluating GPTs' capabilities across various chemistry-related tasks has shown that including contextually similar examples results in superior outcomes compared to approaches that use no example or employ random sampling; The performance of these models improves progressively with the inclusion of additional examples [48, [61, [62]. ICL also boosts the accuracy in more complex regression tasks, rendering GPT-4 competitively effective compared to dedicated machine learning models [63, [64]. Lastly, instead of using specific examples, enriching the context with related information—such as disease backgrounds and synonyms in a fact check task on drug-disease associations [47] —also augments response accuracy. These examples, with in-context learning and context enrichment, underscore the critical role of domain-knowledge in improving the quality and reliability of GPTs' responses in drug discovery tasks.

## 6.3. INSTRUCTION FINETUNING

Task-tuning language models for specific tasks within drug discovery has shown considerable promise, as evidenced by two recent projects. ChatMol [65] is a chatbot based on the T5 model [66], finetuned with experimental property data and molecular spatial knowledge to improve its capabilities in describing and editing target molecules. Task-tuning GPT-3 has demonstrated notable advantages over traditional machine learning approaches, particularly in tasks where training data is small [62]. Task-tuning also significantly improves GPT-3 in extracting DDI triplets, showcasing a substantial F1 score enhancement over GPT-4 with few-shots [67]. These projects demonstrate that task-tuning of foundation models can effectively capture the complex knowledge at the molecule level relevant to drug discovery.

Instruction tuning diverges from task tuning by training an LLM across a spectrum of tasks using instruction-output pairs and enables the model to address new, unseen tasks [68]. DrugAssist [59], a Llama-2-7B-based model, after instruction-tuned with data with individual molecule properties, achieved competitive results when simultaneously optimizing multiple properties. Similarly, DrugChat [58], a Vicuna-13b-based model instruction-tuned with examples from databases like ChEMBL and PubChem, effectively answered open-ended questions about graph-represented drug compounds. Mol-Instructions [69], a large-scale instruction dataset tailored for the biomolecular domain, demonstrated its effectiveness in finetuning models like Llama-7B on a variety of tasks, including molecular property prediction and biomedical text mining.

Task-tuning may be combined with instruction tuning to synergize the strength of each. ChemDFM [70], pre-trained on LLaMa-13B with a chemically rich corpus and further enhanced through instruction tuning, exceled in a range of chemical tasks, particularly in molecular property prediction and reaction prediction, outperforming models like GPT-4 with in-context learning. InstructMol [71] is a multi-modality instruction-tuning-based LLM, featured by a two-stage tuning process, first by instruction tuning with molecule graph-text caption pairs to integrate molecule knowledge and then by task-specific tuning for three drug discovery-related molecular tasks. Applied to Vicuna-7B, InstructMol surpassed other leading open-source LLMs and narrows the performance gap with specialized models [71]. These developments underscore the effectiveness of both task and instruction tuning as strategies for enhancing generalized foundation models with domain-specific knowledge to address specific challenges in drug discovery.

It is important to note that the significant improvements observed through task-tuning and/or instruction-tuning primarily involve open-sourced large language models. These techniques have shown great promise in enhancing model performance in various drug discovery tasks. We noticed that fine-tuning of GPT-3.5 is still in its infancy but encouraging preliminary results have been recently documented in chemical text mining [72]. Unlike its predecessors, GPT-4's fine-tuning capabilities are currently under exploration in an experimental program by OpenAI. As these options become more broadly available, they are expected to significantly advance the field of drug discovery through task/instruction fine-tuning.

## 7. BIOMEDICAL IMAGE UNDERSTANDING

In recent advancements, multimodal AI models have garnered significant attention in biomedical research [73]. Released in late September 2023, GPT-4V(ision) has been the subject of numerous studies that explored its application in image-related tasks across various biomedical topics [74, [75, [76, [77, [78, [79, [80]. For biomedical images, GPT-4V exhibits a performance rivaling professionals in Medical Visual Question Answering [78, [79] and rivals traditional image models in biomedical image classification [81]. For scientific figures, GPT-4V can proficiently explain various plot types and apply domain knowledge to enrich interpretations [82].

Despite the impressive performance, current evaluations reveal significant limitations. OpenAI acknowledges the limitation of GPT-4V in differentiating closely located text and making factual errors in an authoritative tone [83]. The model is not competent in perceiving visual patterns' colors, quantities, and spatial relationships in bioinformatics scientific figures [82]. Image interpretation with domain knowledge from GPT-4V may risk "confirmation bias" [84]: either the observation or conclusion is incorrect, but the supporting knowledge is valid by itself in other irrelevant context [82], or the observation or conclusion is correct, but the supporting knowledge is invalid/irrelevant [85]. Such biases are particularly concerning as users without requisite expertise might be easily misled by these plausible responses.

Prompt engineering has been instrumental in enhancing AI responses to text inputs. The emergence of GPT-4V emphasizes the need to develop equivalent methodologies for visual inputs to refine chatbots' comprehension across modalities. The field of computer vision has already witnessed some progress in this direction [86]. Yang, Z., *et al.* [87] proposes visual referring prompting (VRP) by setting visual pointer references through directly editing input images to augment textual prompts with visual cues. VRP has proven effective in preliminary case studies, leading to the creation of a benchmark like VRPTEST [88] to evaluate its efficacy. Yet, a thorough, quantitative assessment of VRP's impact on GPT-4V's understanding of biomedical images remains to be explored.

## 8. BIOINFORMATICS PROGRAMMING

ChatGPT enables scientists who may not possess advanced programming skills to perform bioinformatics analysis. Users can articulate data characteristics, analysis details, and objectives in natural language, prompting ChatGPT to respond with executable code. In this context, we define “prompt bioinformatics”: the use of natural language instructions (prompts) to guide chatbots for reliable and reproducible bioinformatics data analysis through code generation [13]. This concept differs from the development of bioinformatics chatbot before the GPT era, such as DrBioRight [89] and RiboChat [90]. In prompt bioinformatics, the code is generated on the fly by the chatbot in response to a data analysis description. In addition, the generated code inherently varies across different chat sessions even for the same instruction, adding challenges to new method developments for result reproducibility. Lastly, the concept covers a broad range of bioinformatics topics, particularly those in applied bioinformatics, where data analysis methods are relatively mature.

Early case studies showcase ChatGPT's versatility in addressing diverse bioinformatics coding tasks, from aligning sequencing reads to constructing evolutionary trees [10], and excelling in introductory course exercises [11]. ChatGPT excels at writing short scripts that call existing functions with specific instructions. However, it shows limitations in writing longer, workable code for more complex data analysis with errors often requiring domain-specific knowledge to spot for correction [91].

### 8.1. APPLICATION IN APPLIED BIOINFORMATICS

In applied bioinformatics, established methods for data analysis are prevalent used, enhancing the likelihood of their incorporation into LLM training datasets. Thus, applied bioinformatics emerges as a fertile ground for practicing prompt bioinformatics and evaluating its effectiveness. AutoBA [12], a Python package powered by LLMs, streamlined applied bioinformatics for multi-omics data analysis by autonomously designing analysis plans, generating code, managing package installations, and executing the code. Through testing across 40 varied sequencing-based analysis scenarios, AutoBA with GPT-4 attained a 65% success rate in end-to-end automation [12]. Error message feedback for code correction significantly enhanced this success rate. In addition, AutoBA utilizes retrieval-augmented generation to increase robustness of code generation [12].

Mergen [92] is an R package that automates data analysis through LLM utilization. It crafts, executes, and refines code based on user-provided textual descriptions. The inclusion of file headers in prompts and error message feedback notably improves coding efficacy. The evaluation tasks for Mergen, while relevant to bioinformatics, cater to a general-purpose scope, covering machine learning, statistics, visualization, and data wrangling. Interestingly, the adoption of role-playing does not yield significant enhancements [92], possibly due to the general nature of the tasks and the mismatch between the assumed bioinformatician role and the task requirements.

LLMs exhibit inherent limitations in coding with tools beyond their training datasets. Bioinformaticians typically consult user manuals and source code to master new tools, a process LLMs could emulate. The BioMANIA framework [93] exemplifies this approach by creating conversational chatbots for open-source, well-documented Python tools. By understanding APIs from source code and user manuals, it employs

GPT-4 to generate instructions for API usage. These instructions inform a BERT-based model to suggest top appropriate APIs based on a user's query, with GPT-4 predicting parameters and executing API calls. Evaluation of the method identifies areas for improvement, such as tutorial documentation and API design, guiding the future development of chatbot-compatible tools [93].

## 8.2. BIOMEDICAL DATABASE ACCESS

Structured Query Language (SQL) serves as a pivotal tool for navigating bioinformatics databases. Mastering SQL requires users to have both programming skills and a deep understanding of the database's data schema—prerequisites that many biomedical scientists find challenging. Recent advancements have seen LLM-chatbots like ChatGPT stepping in to translate natural language questions into SQL queries [94], significantly easing database access for non-programmers.

The work by Sima, A.-C. and de Farias, T. M. [95] explored ChatGPT-4's ability to explain and generate SPARQL queries for public biological and bioinformatics databases. Faced with explaining a complex SPARQL query that identifies human genes linked to cancer and their orthologs in rat brains—requiring to combine data from Uniprot, OMA, and Bgee databases—ChatGPT adeptly breaked down the query's elements. However, its attempt to craft a SPARQL query from a natural language description for the same database search revealed inaccuracies that require specific human feedback for correction. Notably, prompts augmented with sematic clues such as variable names and inline comments indicate a substantial improvement in the performance on translating questions into corresponding SPARQL queries, when evaluated on a fine-tuned OpenLlama LLM [96].

Another work by Chen, C. and Stadler, T. [97] applied GPT-3.5 and GPT-4 to convert user inputs into SQL queries for accessing a database of SARS-CoV-2 genomes and their annotations. Through systematic prompting and learning from numerous examples, the chatbot shows proficiency in understanding the database structure and generates accurate queries for 90.6% and 75.2% of the requests with GPT-4 and GPT-3.5, respectively. In addition, the chatbot initiates a new session to explain each query for the users to cross-ref with their own inputs to minimize risks of misunderstandings.

## 8.3. ONLINE TOOLS FOR CODING WITH CHATGPT

Shortly after the release of ChatGPT in November 2022, RTutor.AI emerged as a pioneering web-server powered by the GPT technology dedicated to data analysis. This R-based platform simplifies the process for users to upload a single tabular dataset and articulate their data analysis requirements in natural language. RTutor.AI proficiently manages data importing and type conversion, subsequently leveraging OpenAI's API for R code generation. It executes the generated code and produces downloadable HTML reports including figure plots. A subsequent application, Chatlize.AI, developed by the same team, adopts the tree-of-thought methodology [42] to enhance data analysis exploration. This approach, extending to Python, enables the generation of multiple code versions for a given analysis task, their execution, and comprehensive documentation of the results. Users benefit from the flexibility to select a specific code for further analysis. This feature is particularly valuable for exploratory data analysis, making Chatlize.AI a flexible solution for practicing prompt bioinformatics.

The Code Interpreter, officially integrated into ChatGPT-4 during the summer of 2023 and became a default option in GPT-4o in May 2024, represents a significant advancement in streamlining computational tasks. This feature facilitates a wide array of operations, including data upload, specification of analysis requirements, generation and execution of Python code, visualization of results, and data download, all through natural language instructions. It stands out for its ability to dynamically adapt code in response to runtime errors and self-assess the outcomes of code execution. Despite its broad applicability for general-purpose tasks such as data manipulation and visualization, its utility in bioinformatics data analysis encounters limitations such as the absence of bioinformatics-specific packages and the inability to access external databases [98].

### 8.4. BENCHMARKS FOR BIOINFORMATICS CODING

A thorough assessment of bioinformatics necessitates the establishment of comprehensive benchmarks to cover a broad range of topics in the field. Writing individual functions is a fundamental skill in the development of advanced bioinformatics algorithms. BIOCODER [99] is a benchmark to evaluate language models' proficiency in function writing. This benchmark encompasses over 2,200 Python and Java functions derived from authentic bioinformatics codebases, in addition to 253 functions sourced from the Rosalind project. Comparative analyses have shown that GPT-3.5 and GPT-4 significantly outperform smaller, coding-specific language models on this benchmark. Interestingly, integrating topic-specific context, such as imported objects, into the baseline task descriptions markedly enhances accuracy. However, even the most adept models, namely the GPT series, reach an accuracy ceiling at 60% for GPT-4. A significant proportion of the failures are attributed to syntax or runtime errors [99], suggesting that ChatGPT's effectiveness in bioinformatics coding can be further enhanced through human feedback on error messages.

Execution success is crucial, yet it represents only one facet of evaluating bioinformatics code quality. Sarwal, V., *et al.* [100] proposed a comprehensive evaluation framework that encompassed seven metrics, assessing both subjective and objective dimensions of code writing. These dimensions include readability, correctness, efficiency, simplicity, error handling, code examples, and clarity of input/output specifications. Each metric is scaled from 1 to 10 and normalized independently post-evaluation across models. When applied to a variety of common bioinformatics tasks, this framework highlighted GPT-4's superior performance over alternatives such as BARD and LLaMA. However, the current evaluation remains narrowly focused on a limited number of tasks [100]. Expanding these evaluations to encompass a broader range of bioinformatics domains asks for community-led efforts for a comprehensive appraisal of these language models.

## 9. **CHATBOTS IN BIOINFORMATICS EDUCATION**

The potential of integrating LLMs into bioinformatics education has attracted significant discussions. ChatGPT-3.5 achieves impressive performance in addressing Python programming exercises in an entry-level bioinformatics course [11]. Beyond mere code generation, the utility of chatbots extends to proposing analysis plans, enhancing code readability, elucidating error messages, and facilitating language translation in coding tasks [101]. The effectiveness of a chatbot's response depends on the precision of human instructions, or prompts. In this context, Shue et al. [10] introduced the OPTIMAL model, a framework for prompt refinement through iterative interactions with a chatbot, mirroring the learning curve of bioinformatics beginners assisted by such technologies. To navigate this evolving educational landscape, it becomes imperative to establish guidelines that enable students to critically assess outcomes and articulate constructive feedback to the chatbot for code improvement. Error messages, as one form of such feedback, turn out to be an effective way to boost the coding efficiency of ChatGPT across various studies [10, [12, [92].

The convenience of using chatbots for coding exercises poses a risk of fostering AI overreliance, which will lead to a superficial understanding of the underlying concepts [11, [13, [102]. This AI reliance could undermine students' performance in summative assessments [11]. Innovative evaluation strategies, such as generating multiple-choice questions from student-submitted code to gauge their understanding [103], are needed to counteract this challenge. Such methodologies should aim to deepen students' grasp of the material, ensuring their in-depth understanding of coding concepts.

The art of crafting effective prompts emerges as a critical skill that complements traditional programming competencies. General guidelines are well summarized in a recent commentary [104]. In the context of bioinformatics tasks, these include breaking down a complex task into sub-tasks, enriching context with details (e.g., spelling out package names in code-generation tasks and tissue names for cell type annotation in scRNA-Seq analysis), illustrating intent through examples (e.g., supplying a volcano plot for data visualization task in differentially expressed gene analysis), specifying the output format to facilitate downstream data process while mining gene relationships from literature abstracts, etc. It is important to note that effective prompting is not formulaic. Like coding in bioinformatics and experimental skills for bench works, experience is gained through repetitive experiments [104]. Intriguingly, feedback from a pilot study involving graduate students interacting with ChatGPT for coding highlights the challenges in generating impactful prompts [105]. This prompt-related psychological strain may discourage students from using the chatbot [13]. In this context, the development of a repository featuring carefully crafted prompts for specific bioinformatics analyses—accompanied by quality metrics, reference code, and outcomes—could serve as a valuable resource for students to learn bioinformatics and biomedical informatics aided through prompting with chatbots [10, [13].

In conclusion, while chatbots demonstrate potential as educational tools, their efficacy and effectiveness have not yet been systematically evaluated in classroom settings with controlled experiments. The use of chatbots should be viewed as supplementary to traditional education methodologies [10, [11, [13]. Meanwhile, new assessment methodologies are needed to measure the pedagogical value of chatbots in enhancing bioinformatics learning without diminishing the depth of understanding of concepts and analytical skills.

## 10. **DISCUSSION AND FUTURE PERSPECTIVES**

The year 2023 marked significant progress in leveraging ChatGPT for bioinformatics and biomedical informatics. Early studies affirming its capability in drafting workable code for basic bioinformatics data analysis [10, [11]. The chatbot has also demonstrated competitiveness with SOTA models in other bioinformatics areas, including identifying cell type from single-cell RNA-Seq data [106], performing question-answering tasks in biomedical text mining [107], and generating molecular captions in drug discovery [48]. These achievements underscore ChatGPT’s proficiency in text-generative tasks. Meanwhile, other LLMs are catching up. For example, Google developed Gemini and open-source LLM Gemma, which delivered impressive performance in various tasks. Although their applications in bioinformatics and medical informatics have not been reported, their potentials provide users a viable alternative to ChatGPT.

Current chatbots exhibit limitation in performing biomedical tasks that require reasoning and quantitative analysis, such as regression and classification, as evidenced by references [27, [29, [63, [64, [100]. Though not yet widely adapted in bioinformatics [72], OpenAI’s fine-tuning APIs such as for GPT-3.5 and GPT-4 hold great potential for performance improvements when the training dataset is large. Nevertheless, the accuracy of ChatGPT's responses can be significantly improved through a strategic design of its input instructions with prompt engineering. Incorporating examples into prompts and employing CoT reasoning has proven an effective strategy, as evidenced in various bioinformatics applications [32, [41, [57, [63, [64, [97]. While examples in prompts are sometimes hardcoded, they can also be dynamically and strategically sourced from external knowledge bases or knowledge graphs [57, [59, [61, [108]. This approach, known as retrieval-augmented generation, improves ChatGPT's reliability by sourcing facts from domain-specific knowledge and represents a promising avenue for future development in bioinformatics with chatbots.

Another significant limitation of ChatGPT, like all other LLMs, is hallucination [39, [40]. This occurs when ChatGPT fabricates non-factual content. Instances in bioinformatics applications include inventing functions that do not exist in coding [10], generating false positives when mining gene relationships from biomedical text [41], and fabricating molecular function for gene annotation [36]. While hallucination in code-generation related tasks may be detected through code-execution and partially corrected through error-message feedback, other types require expert knowledge, posing significant risks to general users. To reduce hallucination, one can condition the chatbot with relevant context, such as through RAG, or supplement it with external tools such as task-specific APIs [18]. Despite these strategies, developing evaluation and remediation techniques for detecting hallucinations in LLMs such as ChatGPT —with the accuracy of human experts and the efficiency of computational programs —is urgently needed and remains an ongoing challenge for bioinformatics applications with chatbots.

In this rapidly evolving domain, ChatGPT has experienced several significant upgrades within its first year alone. We acknowledge that not every upgrade enhances performance across the board [109]. Consequently, prompts that are highly effective with the current version for specific tasks may not maintain the same level of efficacy following future updates. The technique of prompt engineering, which includes strategies like role prompting and in-context learning, offers a way to partially counteract this variability [41]. An innovative approach, rather than manually adjusting the prompts, involves instructing ChatGPT to autonomously optimize prompts to align with its latest model iteration. This strategy has shown promise in tasks such as mining gene relationships [41] but remains largely unexplored in other bioinformatics topics and therefore warrants further exploration to fully leverage ChatGPT's capabilities in the field.

Numerous studies repeatedly show that using ChatGPT with human augmentations significantly improve the performance. Iterative human-AI communication plays a pivotal role in this process, where feedback from human operator grounds the chatbot's responses for improved accuracy. This human-in-the-loop methodology is particularly evident in prompt optimization [10] and molecular optimization [56, [59]. For code generation tasks, runtime error message represents commonly used feedback that has been automated into several GPT-based tools [12, [92, [98]. Conversely, the chatbot can also be instructed to provide feedback to human operators. As demonstrated by Chen, C. and Stadler, T. [97], ChatGPT can produce textual descriptions for the generated code through an inverse generation process. Comparing these descriptions with the original instructions from the human operator ensures that the chatbot's output aligns closely with the intended task requirements. This iterative exchange of feedback between AI and human operators enhances the overall quality of the bioinformatics tasks being addressed.

The assessment of ChatGPT's capabilities across various bioinformatics tasks has illuminated both its strengths and weaknesses. Importantly, the reliability of these evaluations largely hinges on the quality of the benchmarks used and the methodologies applied in these assessments. Currently, many benchmarks are available for biomedical text mining and chemistry-related tasks. The development of benchmarks designed specifically for assessing ChatGPT's capability in other bioinformatics tasks, including multimodality, is still in its infancy. It's important to recognize that in generative tasks like coding, producing expected results is not the sole criterion for gauging effectiveness and efficiency. Factors such as the readability of the code and the inclusion of code examples also play crucial roles [100]. Similarly, on prediction or classification tasks, an extension of the evaluation to inspect the text explanations behind the prediction/classification is equally important, as this will facilitate the detection of hidden flaws [85]. Nonetheless, conducting such comprehensive evaluations can be resource-intensive, underscoring the need for community efforts. While alternatives exist for automation, such as transforming tasks into multiple-choice questions or verifying responses against reference texts, for example through lexical overlap or semantic similarity, each method comes with its own set of limitations [7]. Consequently, there is a pressing need to develop new, scalable, and accurate evaluation metrics and benchmark datasets that can accommodate a wide range of bioinformatics tasks, ensuring that assessments are both meaningful and reflective of real-world and cutting-edge applicability.

While aiming for comprehensiveness, our review does not encompass areas that, although outside the direct scope of bioinformatics and biomedical informatics, are closely related and significant. These areas include the management of electronic health records [110, [111], emotion analysis through social media [112], and medical consultation [113, [114]. To mitigate transparency and security concerns, fine-tuning open-source language models deployed locally with task-specific fine-tuning presents a practical approach. Our review has spotlighted such advancements for drug discovery. However, we refer our readers to additional reviews for an expansive understanding of similar developments in other bioinformatics topics, as well as the ethical and legal issues involved [7, [8, [9, [115, [116]. Looking ahead, we envision a future where both online proprietary models such as ChatGPT and open-source, locally deployable finetuned language models co-exist for bioinformatics and biomedical informatics, ensuring users with the most suitable tools to address their specific needs.

## AUTHOR CONTRIBUTIONS

**Gangqing Hu**: Conceptualization, Writing - original draft, Writing - review & editing, Supervision. **Jinge Wang**: Writing - original draft, Writing - review & editing. **Zien Cheng**: Writing - original draft, Writing - review & editing. **Qiuming Yao**: Writing - review & editing. **Li Liu**: Writing - review & editing. **Dong Xu**: Writing - review & editing. All authors have read and approved the final manuscript.

## ACKNOWLEDGEMENTS

This work was partially supported by NIH-NIGMS grants P20 GM103434 and U54 GM-104942, as well as NSF 2125872 (GH). NIH-NLM grant No. R01LM013438 and NIDDK grant No. T32 DK137525 to LL. NLM grant R01LM013392 to DX. The content is solely the responsibility of the authors and does not necessarily represent the official views of the NIH and NSF. The writing was polished by ChatGPT. We thank the following scholars for making comments on the manuscript: Tarcisio Mendes de Farias from Swiss Institute of Bioinformatics (Switzerland), Juexiao Zhou and Xin Gao from King Abdullah University of Science and Technology (Kingdom of Saudi Arabia), and Tanja Stadler from ETH Zürich (Switzerland).

## CONFLICT OF INTEREST STATEMENT

The authors declared no conflict of interest or financial conflicts to disclose.

## ETHICS STATEMENT

There was no sample from human subjects or animals collected for this work.

## FIGURE LEGENDS

**Figure 1: Areas Explored in this Review for ChatGPT's Use in Bioinformatics and Biomedical Informatics in its Year One.**

**Figure 2: ChatGPT-Powered Cell Type Annotation for scRNA-Seq Data Analysis**. In this application, marker genes for each cell cluster are identified using standard pipelines such as Seurat. These markers, along with the corresponding tissue name, are then incorporated into a prompt template, slightly modified

from the GPTCelltype tool [16]. The prompts are submitted to ChatGPT to predict the cell type for each cluster.

**Figure 3: Key Themes from the Application of GPTs and Other LLMs in Drug Discovery Tasks**. The human-in-the-loop section highlights a case study and three interactive tools that facilitate communication between users and chatbots. The in-context learning section emphasizes the use of ad-hoc examples or examples sourced by retrieval-augmented generation to guide chatbots for better performance. The fine-tuning section demonstrates examples on task and/or instruction tuning, primarily with open large language models. Works focusing on the use of GPTs are highlighted in red.

**Figure 4: Illustration of ChatDrug for Conversational Drug Editing with GPT.** In ChatDrug [57], initial prompts are derived from a Prompt Design for Domain-Specific (PDDS) module, which provides tailored templates for specific drug editing tasks. If the response from the chatbot (using GPT-4 as an example) is unsatisfactory, a Retrieval and Domain Feedback (ReDF) module leverages domain knowledge to refine the prompts. Sample prompts, shown in red boxes, are extracted from Liu, S., et al. [57] for a small molecule editing task. In this case, the initial prompts did not yield satisfactory responses (first try), prompting updates from the ReDF module, which subsequently led to satisfactory outcomes (second try).

## Table LEGENDS

**Table 1. Terminologies cited in this review.**

## SUPPLEMENTARY MATERIALS

**Supplementary Table S1: List of studies discussed in this review.**

**Supplementary Table S2: Performance comparison of ChatGPT to baseline models on biomedical text mining tasks.**

**Supplementary Table S3: Performance comparison of ChatGPT to baseline models on drug discovery tasks.**

**Figure 1**

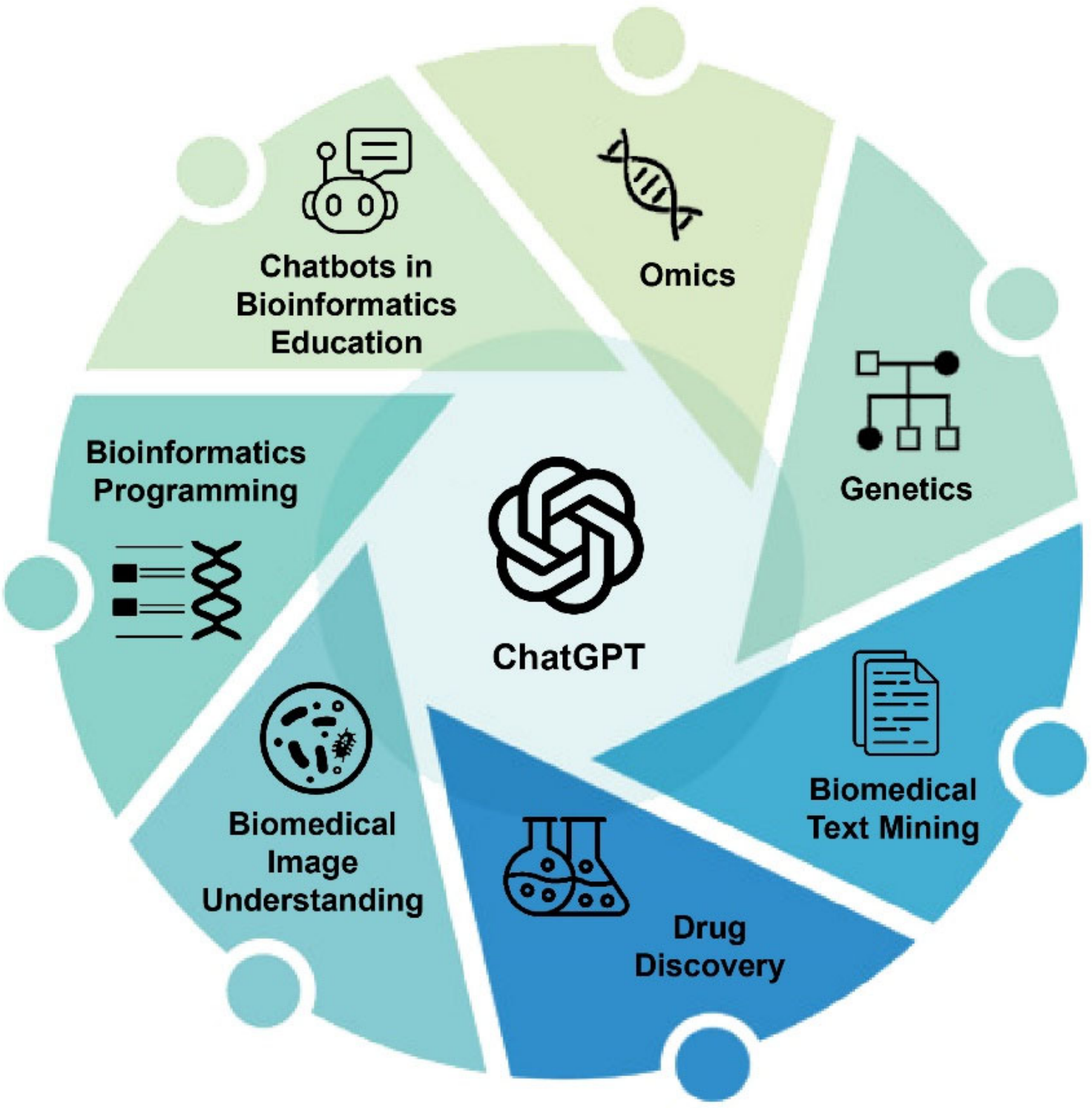

**Figure 2**

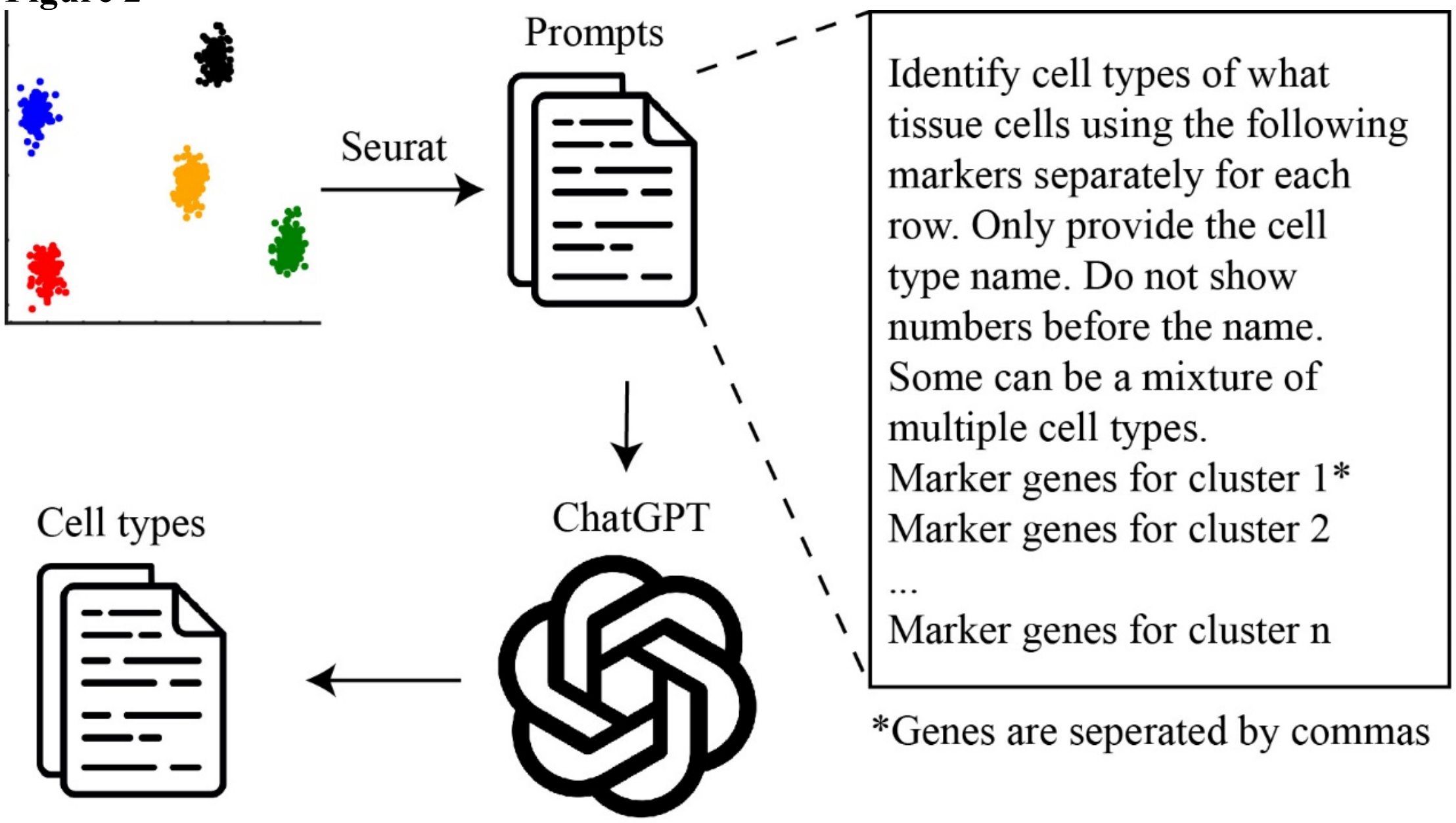
Prompts
Seurat
Identify cell types of what tissue cells using the following markers separately for each row. Only provide the cell type name. Do not show numbers before the name. Some can be a mixture of multiple cell types.
Marker genes for cluster 1*
Marker genes for cluster 2
...
Marker genes for cluster n
*Genes are seperated by commas
ChatGPT
Cell types

**Figure 3**

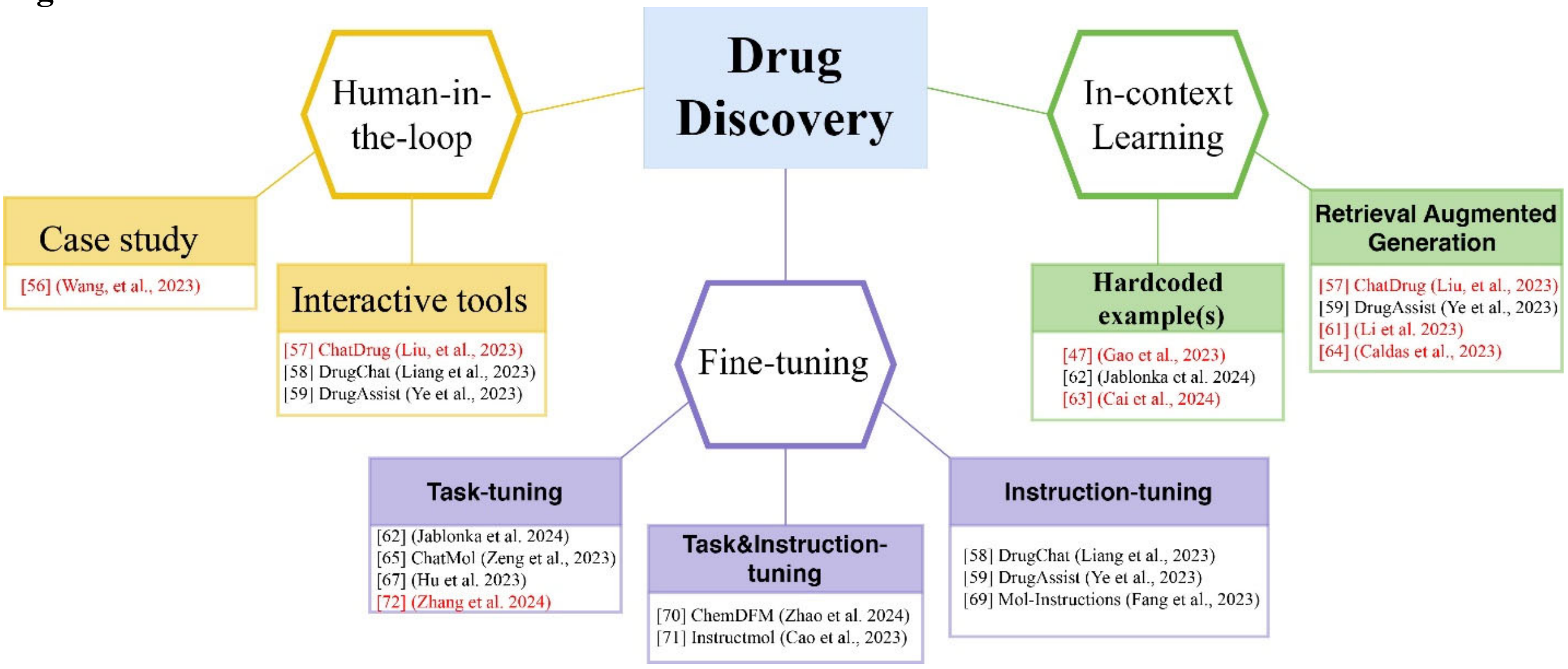

Drug Discovery
Human-in-the-loop
Case study
[56] (Wang, et al., 2023)
Interactive tools
[57] ChatDrug (Liu, et al., 2023)
[58] DrugChat (Liang et al., 2023)
[59] DrugAssist (Ye et al., 2023)
In-context Learning
Hardcoded example(s)
[47] (Gao et al., 2023)
[62] (Jablonka et al. 2024)
[63] (Cai et al., 2024)
Retrieval Augmented Generation
[57] ChatDrug (Liu, et al., 2023)
[59] DrugAssist (Ye et al., 2023)
[61] (Li et al. 2023)
[64] (Caldas et al., 2023)
Fine-tuning
Task-tuning
[62] (Jablonka et al. 2024)
[65] ChatMol (Zeng et al., 2023)
[67] (Hu et al. 2023)
[72] (Zhang et al. 2024)
Task&Instruction-tuning
[70] ChemDFM (Zhao et al. 2024)
[71] Instructmol (Cao et al., 2023)
Instruction-tuning
[58] DrugChat (Liang et al., 2023)
[59] DrugAssist (Ye et al., 2023)
[69] Mol-Instructions (Fang et al., 2023)

**Figure 4**

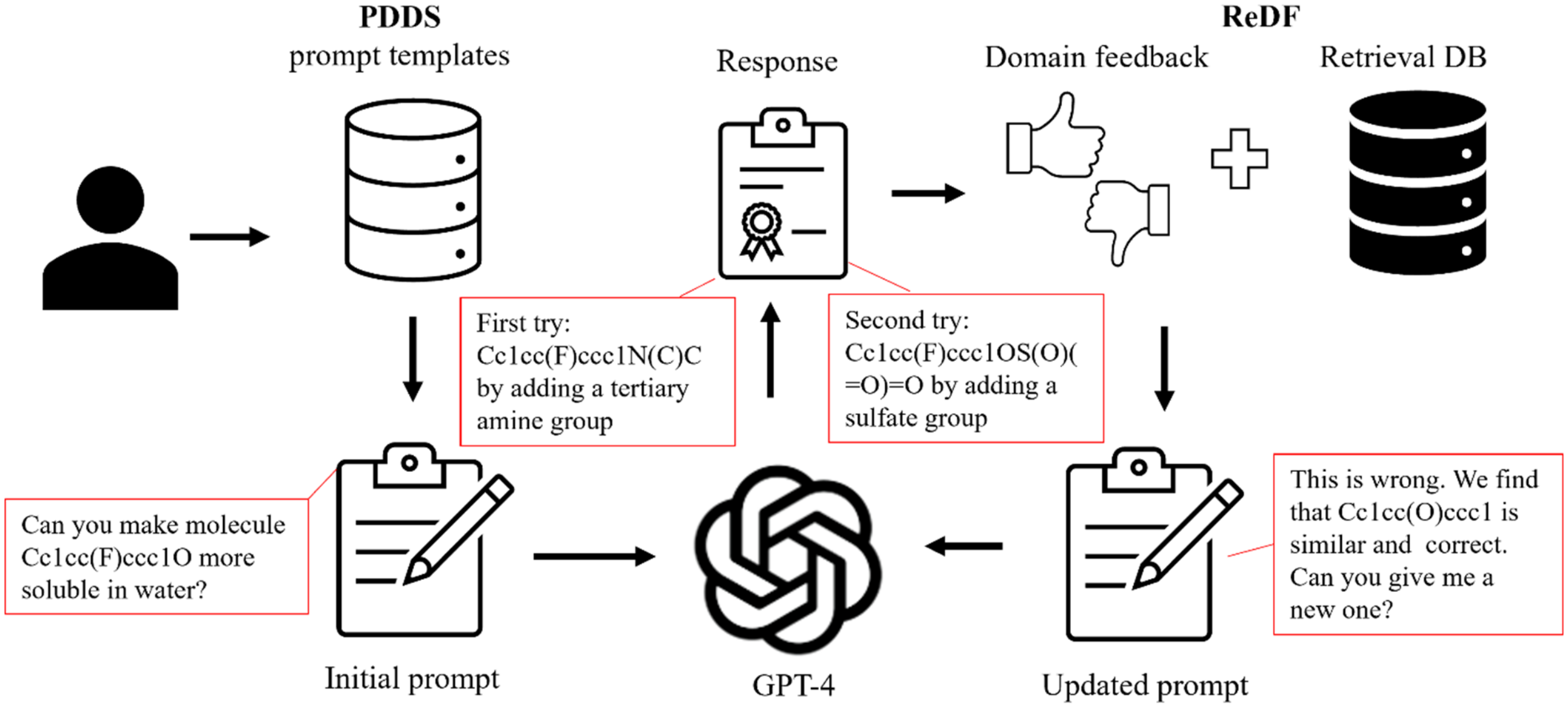

**Table 1.** Terminologies cited in this review.

| **Term** | **Definition** |
|---|---|
| Prompt engineering | The practice of designing and refining input prompts (natural language instruction) to elicit desired responses from a language model chatbot. |
| Zero-shot | A way of prompting where instruction to the chatbot contains no example of a specified task. |
| One-shot | A way of prompting where instruction to the chatbot contains one example of a specified task. |
| Few-shot | A way of prompting where instruction to the chatbot contains more than one examples of that task. |
| Chain of Thought (CoT) | A way of prompting asking the chatbot to think step by step. This approach helps in enhancing the model's ability to solve complex problems by breaking them down into simpler, sequential steps.<br>For one/few-shot, if an example includes details of step-by-step reasoning, the example is called CoT example. |
| Tree of Thought (ToT) | An extension of the Chain of Thought approach, where the model generates a tree-like structure of reasoning steps instead of a linear chain. |
| In-Context Learning (ICL) | A learning paradigm where a model leverages the context provided within the input to adapt and respond to new tasks or information without explicit retraining. |
| Retrieval-Augmented Generation (RAG) | A technique that combines a retriever model, which fetches relevant documents or data, with a generator model, which uses the retrieved information to generate responses or complete tasks. This approach is useful for tasks that require external knowledge or context. |
| Fine-tuning | The process of further training a pre-trained model on a specific dataset or task to improve its performance in that area. |
| Instruction tuning | The process of fine-tuning a pre-trained model to better understand and follow natural language instructions, improving its applicability across different tasks. |
| Task tuning | The process of fine-tuning a pre-trained model on a specific task to enhance its performance on that task. |
| AI hallucination | The phenomenon where a generative AI model produces false or misleading information not supported by the input data or its training. |

# Supplementary Table S1: List of studies discussed in this review.

| Areas | Manuscript Title | Preprint | Journal/Conference | Date of online release | Github | Category |
|---|---|---|---|---|---|---|
| Bioinformatics programming | On the Potential of Artificial Intelligence Chatbots for Data Exploration of Federated Bioinformatics Knowledge Graphs | Arxiv | SeWebMeDa'23: 6th Workshop on Semantic Web solutions for large-scale biomedical data analytics | April 20, 2023 | NA | Evaluation |
| Bioinformatics programming | AI chatbots can boost scientific coding | NA | Nature Ecology & Evolution | April 26, 2023 | NA | Evaluation |
| Bioinformatics programming | Code interpreter for bioinformatics: are we there yet? | NA | Annals of Biomedical Engineering | July 23, 2023 | NA | Evaluation |
| Bioinformatics programming | Biocoder: A benchmark for bioinformatics code generation with contextual pragmatic knowledge | Arxiv | Bioinformatics (in press) | August 31, 2023 | https://github.com/gersteinlab/biocoder | Evaluation |
| Bioinformatics programming | SPARQL Generation: an analysis on fine-tuning OpenLLaMA for Question Answering over a Life Science Knowledge Graph | Arxiv | SWAT4HCLS 2024: The 15th International Conference on Semantic Web Applications and Tools for Health Care and | February 7, 2024 | https://github.com/RIKEN-DKO/Generation_SPARQL | Evaluation |
| Biomedical image understanding | Scientific figures interpreted by chatgpt: Strengths in plot recognition and limits in color perception | BioRxiv | npj Precision Oncology | October 17, 2023 | NA | Evaluation |
| Biomedical image understanding | A pilot study on the efficacy of GPT-4 in providing orthopedic treatment recommendations from MRI reports | NA | Scientific Reports | November 17, 2023 | NA | Evaluation |
| Biomedical text mining | Evaluation of ChatGPT Family of Models for Biomedical Reasoning and Classification | Arxiv | JAMIA | April 5, 2023 | https://github.com/shan23chen/HealthLLM_Eval | Evaluation |
| Biomedical text mining | An extensive benchmark study on biomedical text generation and mining with chatgpt. | BioRxiv | Bioinformatics | April 20, 2023 | NA | Evaluation |
| Biomedical text mining | Opportunities and Challenges for ChatGPT and Large Language Models in Biomedicine and Health | Arxiv | Briefings in Bioinformatics | June 15, 2023 | NA | Evaluation |
| Biomedical text mining | Is ChatGPT a Biomedical Expert?--Exploring the Zero-Shot Performance of Current GPT Models in Biomedical Tasks | Arxiv | 11th BioASQ Workshop at CLEF 2023 | June 28, 2023 | https://github.com/SamyAteia/bioasq | Evaluation |
| Biomedical text mining | A comprehensive evaluation of large language models on benchmark biomedical text processing tasks | Arxiv | Computers in Biology and Medicine journal | October 6, 2023 | https://github.com/tahmedge/llm-eval-biomed | Evaluation |
| Biomedical text mining | Iterative Prompt Refinement for Mining Gene Relationships from ChatGPT | BioRxiv | International Journal of Artificial Intelligence and Robotics Research (in press) | December 23, 2023 | NA | Evaluation |
| Biomedical text mining | EVLncRNAs 3.0: an updated comprehensive database for manually curated functional long non-coding RNAs validated by low-throughput experiments | NA | Nucleic Acids Research | January 1, 2024 | https://www.sdklab-biophysics-dzu.net/EVLncRNAs3/#/ | Application |
| Biomedical text mining | Computational screening of biomarkers and potential drugs for arthrofibrosis based on combination of sequencing and large nature language model | NA | Journal of Orthopaedic Translation | January 20, 2024 | https://chenxi2023.shinyapps.io/afdbv1/ | Application |
| Biomedical text mining | A Comprehensive Evaluation of Large Language Models in Mining Gene Interactions and Pathway Knowledge | BioRxiv | Quantitative Biology (in press) | January 24, 2024 | https://github.com/Muh-aza/LLM | Evaluation |
| Chatbots in bioinformatics education | Evaluating a large language model's ability to solve programming exercises from an introductory bioinformatics course | aRxiv | PLoS Computational Biology | March 7, 2023 | NA | Evaluation |
| Chatbots in bioinformatics education | Empowering Beginners in Bioinformatics with ChatGPT | BioRxiv | Quantitative Biology | March 8, 2023 | NA | Evaluation |
| Drug discovery | The capability of ChatGPT in predicting and explaining common drug-drug interactions | NA | Cureus | March 17, 2023 | NA | Evaluation |
| Drug discovery | ChatGPT-powered Conversational Drug Editing Using Retrieval and Domain Feedback | Arxiv | ICLR 2024 | May 18, 2023 | https://github.com/chao1224/ChatDrug | Application |
| Drug discovery | What can large language models do in chemistry? a comprehensive benchmark on eight tasks | Arxiv | Advances in Neural Information Processing Systems | May 27, 2023 | https://github.com/ChemFoundationModels/ChemLLMBench | Evaluation |
| Drug discovery | Empowering molecule discovery for molecule-caption translation with large language models: A chatgpt perspective | Arxiv | IEEE TRANSACTIONS ON KNOWLEDGEANDDATAENGINEERING (in press) | June 11, 2023 | https://github.com/phenixace/MolReGPT | Evaluation |
| Drug discovery | Mol-instructions: A large-scale biomolecular instruction dataset for large language models | Arxiv | ICLR 2024 | June 13, 2023 | https://github.com/zjunlp/Mol-Instructions | Evaluation |
| Drug discovery | Performance of ChatGPT on the pharmacist licensing examination in Taiwan | NA | Journal of the Chinese Medical Association | July 5, 2023 | NA | Evaluation |
| Drug discovery | Performance of ChatGPT on Chinese national medical licensing examinations: a five-year examination evaluation study for physicians, pharmacists and nurses | medRxiv | BMC Medical Education | August 02, 2023 | NA | Evaluation |
| Drug discovery | Evaluating the performance of ChatGPT in clinical pharmacy: a comparative study of ChatGPT and clinical pharmacists | NA | British journal of clinical pharmacology | August 25, 2023 | NA | Evaluation |
| Drug discovery | Evaluating the sensitivity, specificity, and Accuracy of ChatGPT-3.5, ChatGPT-4, Bing AI, and bard against conventional drug-drug interactions clinical tools | NA | Drug, Healthcare and Patient Safety | September 20, 2023 | NA | Evaluation |
| Drug discovery | Leveraging large language models for predictive chemistry | ChemRxiv | Nature Machine Intelligence | October 17, 2023 | https://github.com/kjappelbaum/gptchem | Evaluation |
| Drug discovery | Examining the Potential of ChatGPT on Biomedical Information Retrieval: Fact-Checking Drug-Disease Associations | NA | Annals of Biomedical Engineering | October 19, 2023 | NA | Evaluation |
| Drug discovery | A generative drug–drug interaction triplets extraction framework based on large language models | NA | Proceedings of the Association for Information Science and Technology | October 22, 2023 | | Evaluation |
| Drug discovery | The potential of GPT-4 as a support tool for pharmacists: analytical study using the Japanese national examination for pharmacists | NA | JMIR Medical Education | October 30, 2023 | NA | Evaluation |
| Drug discovery | Fine-tuning large language models for chemical text mining | ChemRxiv | Chemical Science (in press) | November 16, 2023 | https://github.com/zw-SIMM/SFTChatGPT_for_chemtext_mining | Evaluation |
| Drug discovery | ChatGPT in Drug Discovery: A Case Study on Anticocaine Addiction Drug Development with Chatbots | NA | Journal of Chemical Information and Modeling | November 13, 2023 | https://github.com/wangru25/SGNC | Application |
| Drug discovery | Comprehensive evaluation of molecule property prediction with ChatGPT | NA | Methods | January 17, 2024 | NA | Evaluation |
| Genetics | Analysis of large-language model versus human performance for genetics questions | medRxiv | European Journal of Human Genetics | January 28, 2023 | NA | Evaluation |
| Genetics | Can chatgpt understand genetics? | NA | European Journal of Human Genetics | July 5, 2023 | NA | Evaluation |
| Genetics | Is artificial intelligence getting too much credit in medical genetics? | NA | American Journal of Medical Genetics | August 22, 2023 | NA | Evaluation |
| Genetics | Genetic counselors' utilization of ChatGPT in professional practice: A cross-sectional study | NA | American Journal of Medical Genetics | December 8, 2023 | NA | Evaluation |
| Omics | Assessing GPT-4 for cell type annotation in single-cell RNA-seq analysis | BioRxiv | Nature Methods | April 21, 2023 | https://github.com/Winnie09/GPTCelltype_Paper | Evaluation |
| Omics | Genegpt: Augmenting large language models | Arxiv | Bioinformatics | May 16, 2023 | https://github.com/ncbi/GeneGPT. | Application |
| Bioinformatics programming | GenSpectrum Chat: Data Exploration in Public Health Using Large Language Models | Arxiv | NA | May 23, 2023 | https://cov-spectrum.org/chat | Application |
| Bioinformatics programming | BioMANIA: Simplifying bioinformatics data analysis through conversation | BioRxiv | NA | November 1, 2023 | https://github.com/batmen-lab/BioMANIA | Application |
| Bioinformatics programming | Leveraging large language models for data analysis automation | BioRxiv | NA | December 21, 2023 | https://github.com/BIMSBbioinfo/mergen-manuscript | Application |
| Bioinformatics programming | An AI Agent for Fully Automated Multi-omic Analyses | BioRxiv | NA | January 5, 2024 | https://github.com/JoshuaChou2018/AutoBA | Application |
| Bioinformatics programming | BioLLMBench: A Comprehensive Benchmarking of Large Language Models in Bioinformatics | BioRxiv | NA | January 16, 2024 | NA | Evaluation |
| Biomedical image understanding | Accuracy of a vision-language model on challenging medical cases | Arxiv | NA | November 9, 2023 | NA | Evaluation |
| Biomedical image understanding | Performance of multimodal gpt-4v on usmle with image: Potential for imaging diagnostic support with explanations | medRxiv | NA | November 15, 2023 | NA | Evaluation |
| Biomedical image understanding | GPT-4V exhibits human-like performance in biomedical image classification | BioRxiv | NA | January 1, 2024 | https://github.com/Winnie09/gptimage | Evaluation |
| Biomedical image understanding | Hidden Flaws Behind Expert-Level Accuracy of GPT-4 Vision in Medicine | Arxiv | NA | January 16, 2024 | NA | Evaluation |
| Biomedical text mining | Evaluation of GPT and BERT-based models on identifying protein-protein interactions in biomedical text | Arxiv | NA | March 30, 2023 | NA | Evaluation |
| Biomedical text mining | Large language models in biomedical natural language processing: benchmarks, baselines, and recommendations | Arxiv | NA | May 10, 2023 | https://github.com/qingyu-qc/gpt_bionlp_benchmark | Evaluation |
| Biomedical text mining | PlantConnectome: knowledge networks encompassing> 100,000 plant article abstracts | BioRxiv | NA | July 15, 2023 | https://connectome.plant.tools/ | Application |
| Biomedical text mining | From answers to insights: Unveiling the strengths and limitations of chatgpt and biomedical knowledge graphs | Res Sq | NA | August 1, 2023 | NA | Evaluation |
| Biomedical text mining | ChatGPT usage in the Reactome curation process | BioRxiv | NA | November 8, 2023 | NA | Application |
| Biomedical text mining | reguloGPT: Harnessing GPT for Knowledge Graph Construction of Molecular Regulatory Pathways | BioRxiv | NA | January 30, 2024 | https://github.com/Huang-AI4Medicine-Lab/reguloGPT | Application |
| Drug discovery | Bayesian optimization of catalysts with in-context learning | Arxiv | NA | April 11, 2023 | https://github.com/ur-whitelab/BO-LIFT | Evaluation |
| Drug discovery | DrugChat: towards enabling ChatGPT-like capabilities on drug molecule graphs | Arxiv | NA | May 18, 2023 | https://github.com/UCSD-AI4H/drugchat | Application |
| Drug discovery | Interactive molecular discovery with natural language | Arxiv | NA | June 21, 2023 | https://github.com/Ellenzzn/ChatMol/tree/main | Application |
| Drug discovery | Instructmol: Multi-modal integration for building a versatile and reliable molecular assistant in drug discovery | Arxiv | NA | November 27, 2023 | https://idea-xl.github.io/InstructMol/ | Application |
| Drug discovery | Drugassist: A large language model for molecule optimization | Arxiv | NA | December 28, 2023 | https://github.com/blazerye/DrugAssist | Application |
| Drug discovery | ChemDFM: Dialogue Foundation Model for Chemistry | Arxiv | NA | January 26, 2024 | NA | Application |
| Omics | GeneTuring tests GPT models in genomics | BioRxiv | NA | March 13, 2023 | NA | Evaluation |

For preprints not yet formally published, those cited to support shared findings across independent works are underlined.

# Supplementary Table S2: Performance comparison of ChatGPT to baseline models on biomedical text mining tasks.

| | | | | ChatGPT | | | | | | | | BERT | | | | | | Alternative LLMs to ChatGPT | | | | | | | | | | | | | | | | | | | Others |
|---|---|---|---|---|---|---|---|---|---|---|---|---|---|---|---|---|---|---|---|---|---|---|---|---|---|---|---|---|---|---|---|---|---|---|---|---|---|
| Reference Title | Tasks | Benchmark | Evaluation Metrics | GPT-3.5 | 3.5 (one-shot) | 3.5 (few-shot) | 3.5 (few-shot | GPT-4 | GPT-4 (one-shot) | GPT-4 (few-shot) | (few-shot CoT) | PubMedBERT | BioLinkBERT-Base | kBERT-Large | BioBERT | PubmedBert | SciBERT | Claude-2 | Claude-Instant | Cohere | Claude-Pro | PaLM-2 | Bard | Codellama-34 | Wizardlm-70b | Wizardlm-13b | Falcon-180b | Mistral-7b | Chatglm2-6b | Vicuna-7b | Vicuna-33b | Vicuna-13b | llama-2-70b | LLaMa-2-13b | llama-2-7b | Qwen-14b | BoW |
| An extensive benchmark study on biomedical text generation and mining with chatgpt. | Named entity recognition | BC5-chem | F1 entity-level | 60.3 | - | - | - | - | - | - | - | 93.33 | 93.75 | 94.04 | - | - | - | - | - | - | - | - | - | - | - | - | - | - | - | - | - | - | - | - | - | - | |
| | Named entity recognition | BC5-disease | F1 entity-level | 51.77 | - | - | - | - | - | - | - | 85.62 | 86.1 | 86.39 | - | - | - | - | - | - | - | - | - | - | - | - | - | - | - | - | - | - | - | - | - | - | |
| | Named entity recognition | NCBI-disease | F1 entity-level | 50.49 | - | - | - | - | - | - | - | 87.82 | 88.18 | 88.76 | - | - | - | - | - | - | - | - | - | - | - | - | - | - | - | - | - | - | - | - | - | - | |
| | Named entity recognition | BC2GM | F1 entity-level | 37.54 | - | - | - | - | - | - | - | 84.52 | 84.9 | 85.18 | - | - | - | - | - | - | - | - | - | - | - | - | - | - | - | - | - | - | - | - | - | - | |
| | Named entity recognition | JNLPBA | F1 entity-level | 41.25 | - | - | - | - | - | - | - | 80.06 | 79.03 | 80.06 | - | - | - | - | - | - | - | - | - | - | - | - | - | - | - | - | - | - | - | - | - | - | |
| | PICO extraction | EBM PICO | Macro F1 word-level | 55.59 | - | - | - | - | - | - | - | 73.38 | 73.97 | 74.19 | - | - | - | - | - | - | - | - | - | - | - | - | - | - | - | - | - | - | - | - | - | - | |
| | Relation extraction | ChemProt | Micro F1 | 34.16 | 48.64 | - | - | - | | - | - | 77.24 | 77.57 | 79.98 | - | - | - | - | - | - | - | - | - | - | - | - | - | - | - | - | - | - | - | - | - | - | |
| | Relation extraction | DDI | Micro F1 | 51.62 | - | - | - | - | - | - | - | 82.36 | 82.72 | 83.35 | - | - | - | - | - | - | - | - | - | - | - | - | - | - | - | - | - | - | - | - | - | - | |
| | Relation extraction | GAD | Micro F1 | 52.43 | - | - | - | - | - | - | - | 82.34 | 84.39 | 84.9 | - | - | - | - | - | - | - | - | - | - | - | - | - | - | - | - | - | - | - | - | - | - | |
| | Sentence similarity | BIOSSES | Pearson | 43.75 | - | - | - | - | - | - | - | 92.3 | 93.25 | 93.63 | - | - | - | - | - | - | - | - | - | - | - | - | - | - | - | - | - | - | - | - | - | - | |
| | Document classification | HoC | Average Micro F1 | 51.22 | - | - | - | - | - | - | - | 82.34 | 84.39 | 84.9 | - | - | - | - | - | - | - | - | - | - | - | - | - | - | - | - | - | - | - | - | - | - | |
| | Question answering | PubMedQA | Accuracy | 76.45 | - | - | - | - | - | - | - | 55.84 | 70.2 | 72.18 | - | - | - | - | - | - | - | - | - | - | - | - | - | - | - | - | - | - | - | - | - | - | |
| | Question answering | BioASQ | Accuracy | 88.57 | - | - | - | - | - | - | - | 87.56 | 91.43 | 94.82 | - | - | - | - | - | - | - | - | - | - | - | - | - | - | - | - | - | - | - | - | - | - | |
| Large language models in biomedical natural language processing: benchmarks, baselines, and recommendations | Named entity recognition | BC5CDR-chemical | F1 entity-level | 68.36 | 72.1 | - | - | 81.9 | 82.43 | - | - | 93.5 | - | - | - | - | - | - | - | - | - | - | - | - | - | - | - | - | - | - | - | - | - | - | - | - | |
| | Named entity recognition | NCBI-disease | F1 entity-level | 38.02 | 42.74 | - | - | 57.85 | 58.39 | - | - | 89.86 | - | - | - | - | - | - | - | - | - | - | - | - | - | - | - | - | - | - | - | - | - | - | - | - | |
| | Relation extraction | ChemProt | Micro F1 | 57.43 | 57.71 | - | - | 66.18 | 66.82 | - | - | 78.32 | - | - | - | - | - | - | - | - | - | - | - | - | - | - | - | - | - | - | - | - | - | - | - | - | |
| | Relation extraction | DDI2013 | Micro F1 | 33.49 | 34.34 | - | - | 63.25 | 61.76 | - | - | 80.23 | - | - | - | - | - | - | - | - | - | - | - | - | - | - | - | - | - | - | - | - | - | - | - | - | |
| | Multi-label classification | HoC | Label-wise macro F1 | 65.72 | 69.32 | - | - | 74.74 | 74.02 | - | - | 89.15 | - | - | - | - | - | - | - | - | - | - | - | - | - | - | - | - | - | - | - | - | - | - | - | - | |
| | Multi-label classification | LitCovid | Label-wise macro F1 | 63.9 | 65.31 | - | - | 67.46 | 68.39 | - | - | 87.24 | - | - | - | - | - | - | - | - | - | - | - | - | - | - | - | - | - | - | - | - | - | - | - | - | |
| | Semantic similarity and reasoning | PubMedQA | Pearson | 35.53 | 30.11 | - | - | 43.74 | 53.61 | - | - | 36.76 | - | - | - | - | - | - | - | - | - | - | - | - | - | - | - | - | - | - | - | - | - | - | - | - | |
| | Semantic similarity and reasoning | BIOSSES | Pearson | 87.86 | 91.94 | - | - | 88.32 | 89.22 | - | - | 93.32 | - | - | - | - | - | - | - | - | - | - | - | - | - | - | - | - | - | - | - | - | - | - | - | - | |
| | Text summarization | summarization | ROUGE-1 | 6.08 | 23.2 | - | - | 39.97 | 40.54 | - | - | 44.89 | - | - | - | - | - | - | - | - | - | - | - | - | - | - | - | - | - | - | - | - | - | - | - | - | |
| | Text summarization | MS^2 | ROUGE-1 | 17.31 | 12.11 | - | - | 18.77 | 19.19 | - | - | 20.79 | - | - | - | - | - | - | - | - | - | - | - | - | - | - | - | - | - | - | - | - | - | - | - | - | |
| | Text simplification | Cochrane PLS | Flesch-Kincaid score | 13.0505 | 13.18 | - | - | 12.0001 | 13.12 | - | - | 12.64 | - | - | - | - | - | - | - | - | - | - | - | - | - | - | - | - | - | - | - | - | - | - | - | - | |
| | Text simplification | PLOS text simplification | Flesch-Kincaid score | 14.0605 | 13.92 | - | - | 13.219 | 13.24 | - | - | 14.66 | - | - | - | - | - | - | - | - | - | - | - | - | - | - | - | - | - | - | - | - | - | - | - | - | |
| Evaluation of ChatGPT Family of Models for Biomedical Reasoning and Classification | Classification | "Advice in discussion sections" | Micro F1 | 50.6 | 51.3 | 47.5 | 67.1 | 50.9 | - | - | 64.8 | - | - | - | 80 (100% fine-tuning) | - | - | - | - | - | - | - | - | - | - | - | - | - | - | - | - | - | - | - | - | - | 59.3 (100% fine-tuning) |
| | Classification | "Advice in unstructured abstracts" | Micro F1 | 48.9 | 55.4 | 40.3 | 67 | 47.8 | - | - | 71.2 | - | - | - | 82.1 (100% fine-tuning) | - | - | - | - | - | - | - | - | - | - | - | - | - | - | - | - | - | - | - | - | - | 64 (100% fine-tuning) |
| | Classification | "Advice in structured abstracts" | Micro F1 | 54.8 | 59.3 | 49.5 | 71.8 | 51.7 | - | - | 77 | - | - | - | 90.2 (100% fine-tuning) | - | - | - | - | - | - | - | - | - | - | - | - | - | - | - | - | - | - | - | - | - | 75.3 (100% fine-tuning) |
| | Reasoning | "Causal relation detection" | Micro F1 | 28.8 | 39.6 | 54.2 | 64.9 | 30.1 | - | - | 68.2 | - | - | - | 85.1 (100% fine-tuning) | - | - | - | - | - | - | - | - | - | - | - | - | - | - | - | - | - | - | - | - | - | 67.5 (100% fine-tuning) |
| A comprehensive evaluation of large language models on benchmark biomedical text processing tasks | Relation extraction | BC5CDR | F1 | 43.29 | - | - | - | - | - | - | - | - | - | - | - | - | - | 53.37 | - | - | - | 54.3 | - | - | - | - | - | - | - | - | - | - | - | 53.28 | - | - | - |
| | Relation extraction | KD-DTI | F1 | 29.74 | - | - | - | - | - | - | - | - | - | - | - | - | - | 28.84 | - | - | - | 38.44 | - | - | - | - | - | - | - | - | - | - | - | 24.21 | - | - | - |
| | Relation extraction | DDI | F1 | 46.43 | - | - | - | - | - | - | - | - | - | - | - | - | - | 42.62 | - | - | - | 22.5 | - | - | - | - | - | - | - | - | - | - | - | 24.03 | - | - | - |
| | Text classification | HoC | F1 | 59.26 | - | - | - | - | - | - | - | - | - | - | - | - | - | 34.93 | - | - | - | 61.03 | - | - | - | - | - | - | - | - | - | - | - | 41.82 | - | - | - |
| | Text classification | LitCovid | F1 | 29.63 | - | - | - | - | - | - | - | - | - | - | - | - | - | 7.6 | - | - | - | 37.5 | - | - | - | - | - | - | - | - | - | - | - | 11.34 | - | - | - |
| | Question answering | PubMedQA | Accuracy | 54.4 | - | - | - | - | - | - | - | - | - | - | - | - | - | 57.2 | - | - | - | 59.6 | - | - | - | - | - | - | - | - | - | - | - | 61.4 | - | - | - |
| | Question answering | MediQA-2019 | Accuracy | 73.26 | - | - | - | - | - | - | - | - | - | - | - | - | - | 65.13 | - | - | - | 52.12 | - | - | - | - | - | - | - | - | - | - | - | 56.01 | - | - | - |
| | Entity linking dataset | BC5CDR | Recall@1 | 54.9 | - | - | - | - | - | - | - | - | - | - | - | - | - | 78.01 | - | - | - | 52.14 | - | - | - | - | - | - | - | - | - | - | - | 66.52 | - | - | - |
| | Entity linking dataset | Cometa | Recall@1 | 43.45 | - | - | - | - | - | - | - | - | - | - | - | - | - | 53.29 | - | - | - | 48.76 | - | - | - | - | - | - | - | - | - | - | - | 40.67 | - | - | - |
| | Entity linking dataset | NCBI | Recall@1 | 52.19 | - | - | - | - | - | - | - | - | - | - | - | - | - | 70.21 | - | - | - | 38.44 | - | - | - | - | - | - | - | - | - | - | - | 59.17 | - | - | - |
| | Named entity recognition | BC2GM | F1 | 31.99 | - | - | - | - | - | - | - | - | - | - | - | - | - | 40.45 | - | - | - | 32.75 | - | - | - | - | - | - | - | - | - | - | - | 5.95 | - | - | - |
| | Named entity recognition | BC4CHEMD | F1 | 26.01 | - | - | - | - | - | - | - | - | - | - | - | - | - | 35.18 | - | - | - | 25.83 | - | - | - | - | - | - | - | - | - | - | - | 6.64 | - | - | - |
| | Named entity recognition | BC5CDR-chem | F1 | 41.25 | - | - | - | - | - | - | - | - | - | - | - | - | - | 58.05 | - | - | - | 48.08 | - | - | - | - | - | - | - | - | - | - | - | 12.21 | - | - | - |
| | Named entity recognition | BC5CDR-disease | F1 | 32.26 | - | - | - | - | - | - | - | - | - | - | - | - | - | 50.13 | - | - | - | 33.72 | - | - | - | - | - | - | - | - | - | - | - | 5.68 | - | - | - |
| | Named entity recognition | JNLPBA | F1 | 31.89 | - | - | - | - | - | - | - | - | - | - | - | - | - | 34.62 | - | - | - | 21.18 | - | - | - | - | - | - | - | - | - | - | - | 4.3 | - | - | - |
| | Named entity recognition | NCBI-disease | F1 | 33.39 | - | - | - | - | - | - | - | - | - | - | - | - | - | 45.75 | - | - | - | 31.15 | - | - | - | - | - | - | - | - | - | - | - | 4.58 | - | - | - |
| | Named entity recognition | linnaeus | F1 | 5.14 | - | - | - | - | - | - | - | - | - | - | - | - | - | 13.91 | - | - | - | 6.44 | - | - | - | - | - | - | - | - | - | - | - | 1.42 | - | - | - |
| | Named entity recognition | s800 | F1 | 15.57 | - | - | - | - | - | - | - | - | - | - | - | - | - | 24.07 | - | - | - | 16.96 | - | - | - | - | - | - | - | - | - | - | - | 1.87 | - | - | - |
| | Text summarization | iCliniq | ROUGE-1 | 30.5 | - | - | - | - | - | - | - | - | - | - | - | - | - | 28.8 | - | - | - | 21.9 | - | - | - | - | - | - | - | - | - | - | - | 20 | - | - | - |
| | Text summarization | HealthCareMagic | ROUGE-1 | 28.1 | - | - | - | - | - | - | - | - | - | - | - | - | - | 24.4 | - | - | - | 25.9 | - | - | - | - | - | - | - | - | - | - | - | 16.7 | - | - | - |
| | Text summarization | MeQSum | ROUGE-1 | 30 | - | - | - | - | - | - | - | - | - | - | - | - | - | 31.7 | - | - | - | 31.5 | - | - | - | - | - | - | - | - | - | - | - | 21.2 | - | - | - |
| | Text summarization | MEDIQA-QS | ROUGE-1 | 30.6 | - | - | - | - | - | - | - | - | - | - | - | - | - | 32 | - | - | - | 29.7 | - | - | - | - | - | - | - | - | - | - | - | 23.3 | - | - | - |
| | Text summarization | MEDIQA-MAS | ROUGE-1 | 38.9 | - | - | - | - | - | - | - | - | - | - | - | - | - | 13.4 | - | - | - | 15.3 | - | - | - | - | - | - | - | - | - | - | - | 13.7 | - | - | - |
| | Text summarization | MEDIQA-ANS | ROUGE-1 | 28.7 | - | - | - | - | - | - | - | - | - | - | - | - | - | 28.6 | - | - | - | 25.4 | - | - | - | - | - | - | - | - | - | - | - | 28 | - | - | - |
| | Text summarization | PLOS (Abstract) | ROUGE-1 | 39.65 | - | - | - | - | - | - | - | - | - | - | - | - | - | 42.25 | - | - | - | 25.09 | - | - | - | - | - | - | - | - | - | - | - | 41.78 | - | - | - |
| | Text summarization | Summarization) | ROUGE-1 | 39.13 | - | - | - | - | - | - | - | - | - | - | - | - | - | 36.16 | - | - | - | 30.7 | - | - | - | - | - | - | - | - | - | - | - | 36.33 | - | - | - |
| Evaluation of GPT and BERT-based models on identifying protein-protein interactions in biomedical text | Relation extraction | LLL | F1-Score | 76.72 (with protein dictionary) | - | - | - | 86.49 (with protein dictionary) | - | - | - | - | - | - | 86.84 | 85.42 | 84.66 | - | - | - | - | - | - | - | - | - | - | - | - | - | - | - | - | - | - | - | - |
| | Relation extraction | IEPA | F1-Score | 31.25 (with protein dictionary) | - | - | - | 48.37 (with protein dictionary) | - | - | - | - | - | - | 78.81 | 78.49 | 75.53 | - | - | - | - | - | - | - | - | - | - | - | - | - | - | - | - | - | - | - | - |
| | Relation extraction | HPRD50 | F1-Score | 52.9 (with protein dictionary) | - | - | - | 57.24 (with protein dictionary) | - | - | - | - | - | - | 74.95 | 79.65 | 74.67 | - | - | - | - | - | - | - | - | - | - | - | - | - | - | - | - | - | - | - | - |
| A Comprehensive Evaluation of Large Language Models in Mining Gene Interactions and Pathway Knowledge | Relation extraction | Gene regulatory relations | F1 | 0.1485 | - | - | - | 0.4448 | - | - | - | - | - | - | - | - | - | 0.422 | 0.326 | 0.374 | 0.439 | 0.261 | 0.279 | 0.144 | 0.045 | 0.096 | 0.279 | 0.123 | 0.182 | 0.123 | 0.141 | 0.181 | 0.158 | 0.091 | 0.192 | 0.101 | - |
| | Relation extraction | KEGG pathway recognition | Jaccard similarity | 0.1358 | - | - | - | 0.2778 | - | - | - | - | - | - | - | - | - | 0.189 | 0.152 | 0.209 | 0.266 | 0.105 | 0.192 | 0.162 | 0.135 | 0.146 | 0.224 | 0.143 | 0.088 | 0.174 | 0.166 | 0.154 | 0.154 | 0.171 | 0.221 | 0.084 | - |

Values are extracted from corresponding reference listed in the first column. In red are numbers that are no better than at least one kind of GPT.

# Supplementary Table S3: Performance comparison of ChatGPT to baseline models on drug discovery tasks.

| | | | | ChatGPT | | | Baseline models | | | | | | | | | | | | | | | | | | |
|---|---|---|---|---|---|---|---|---|---|---|---|---|---|---|---|---|---|---|---|---|---|---|---|---|---|
| Reference Title | Tasks | Benchmark | Evaluation Metrics | GPT-3.5 | GPT-3.5 (fine-tuned) | GPT-4 | Davinci-003 | GAL-30B | MolT5-Large | LLama-2-7B | Llama2-13B-chat | BART-base | T5-base | T5-large | MolT5-base | MolT5-large | text-ada-001 (fine-tuned) | Ridge Regression | Neighbor (KNN) | ada embeddings | SolTranNet | SMILES-BERT | MolBERT | Regression Transformer | MolFormer |
| What can large language models do in chemistry? a comprehensive benchmark on eight tasks | Name Prediction ("smiles2formula") | PubChem | Accuracy | 0.052 ('Scaffold,k=20') | - | 0.086 ('Scaffold,k=20') | 0.006 ("Scaffold,k=20") | - | - | - | 0.01 ("Scaffold,k=20") | - | - | - | - | - | - | - | - | - | - | - | - | - | - |
| | Property Prediction | BBBP | F1 | 0.463 ('Scaffold,k=20') | - | 0.587 ('Scaffold,k=20') | 0.378 ('Scaffold,k=20') | 0.074 ('Scaffold,k=20') | - | - | 0.002 ('Scaffold,k=20') | - | - | - | - | - | - | - | - | - | - | - | - | - | - |
| | Property Prediction | HIV | F1 | 0.406 ('Scaffold,k=20') | - | 0.666 ('Scaffold,k=20') | 0.649 ('Scaffold,k=20') | 0.025 ('Scaffold,k=20') | - | - | 0.045 ('Scaffold,k=20') | - | - | - | - | - | - | - | - | - | - | - | - | - | - |
| | Property Prediction | BACE | F1 | 0.807 ('Scaffold,k=20') | - | 0.797 ('Scaffold,k=20') | 0.832 ('Scaffold,k=20') | 0.014 ('Scaffold,k=20') | - | - | 0.069 ('Scaffold,k=20') | - | - | - | - | - | - | - | - | - | - | - | - | - | - |
| | Property Prediction | Tox21 | F1 | 0.529 ('Scaffold,k=20') | - | 0.563 ('Scaffold,k=20') | 0.518 ('Scaffold,k=20') | 0.077 ('Scaffold,k=20') | - | - | 0.047 ('Scaffold,k=20') | - | - | - | - | - | - | - | - | - | - | - | - | - | - |
| | Property Prediction | ClinTox | F1 | 0.369 ('Scaffold,k=20') | - | 0.736 ('Scaffold,k=20') | 0.85 ('Scaffold,k=20') | 0.081 ('Scaffold,k=20') | - | - | 0.001 ('Scaffold,k=20') | - | - | - | - | - | - | - | - | - | - | - | - | - | - |
| | Yield Prediction | Buchwald-Hartwig | Accuracy | 0.585 ('random, k = 8') | - | 0.8 ('random, k = 8') | 0.467 ('random, k = 8') | 0 | - | - | 0.008 | - | - | - | - | - | - | - | - | - | - | - | - | - | - |
| | Yield Prediction | Suzuki-Miyaura | Accuracy | 0.542 ('random, k = 8') | - | 0.764 ('random, k = 8') | 0.341 ('random, k = 8') | 0.008 | - | - | 0.006 | - | - | - | - | - | - | - | - | - | - | - | - | - | - |
| | Reaction Prediction | USPTO-Mixed | Accuracy | 0.184 ('Scaffold, k=20') | - | 0.23 ('Scaffold, k=20') | 0.218 ('Scaffold, k=20') | 0.036 ('Scaffold, k=5') | - | - | 0.032 ('Scaffold, k=20') | - | - | - | - | - | - | - | - | - | - | - | - | - | - |
| | Reagents Selection | Suzuki-Miyaura | Top-1 Accuracy | 0.4 | - | 0.299 | 0.178 | 0.107 | - | - | 0.145 | - | - | - | - | - | - | - | - | - | - | - | - | - | - |
| | Retrosynthesis | USPTO-50k | Top-1Accuracy | 0.022 ('Scaffold, k=20') | - | 0.096 ('Scaffold, k=20') | 0.122 ('Scaffold, k=20') | 0.016 ('Scaffold, k=5') | - | - | 0 ('Scaffold, k=20') | - | - | - | - | - | - | - | - | - | - | - | - | - | - |
| | Text-Based Molecule Design | ChEBI-20 | BLEU | 0.479 ('Scaffold, k=10') | - | 0.816 ('Scaffold, k=10') | 0.741 ('Scaffold, k=10') | 0.004 | 0.601 | - | 0.626 ('Scaffold, k=10') | - | - | - | - | - | - | - | - | - | - | - | - | - | - |
| | Molecule Captioning | ChEBI-20 | BLEU-2 | 0.468 ('Scaffold,k=10') | - | 0.464 ('Scaffold,k=10') | 0.488 ('Scaffold,k=10') | 0.008 | 0.482 | - | 0.197 ('Scaffold,k=10') | - | - | - | - | - | - | - | - | - | - | - | - | - | - |
| Empowering molecule discovery for molecule-caption translation with large language models: A chatgpt perspective | Molecule Captioning | ChEBI-20 | BLEU-2 | 0.565 ('10-shot MolReGPT') | - | 0.607 ('10-shot MolReGPT') | - | - | - | 0.489 (2-shot MolReGPT) | - | - | 0.511 | 0.558 | 0.54 | 0.594 | - | - | - | - | - | - | - | - | - |
| | Text-Based Molecule Design | ChEBI-20 | BLEU | 0.79 ('10-shot MolReGPT') | - | 0.857 ('10-shot MolReGPT') | - | - | - | 0.693 (2-shot MolReGPT) | - | - | 0.762 | 0.854 | 0.769 | 0.854 | - | - | - | - | - | - | - | - | - |
| Bayesian optimization of catalysts with in-context learning | Property Prediction ("aqueous solubility") | ESOL | RMSE (the lower the better) | - | - | 0.773 ('topk') | 1.185 ('topk') | - | - | - | - | - | - | - | - | - | 1.558 ('topk') | - | 2.443 | 2.652 ('topk') | 2.99 | 0.47 | 0.531 | 0.73 | 0.278 |
| | Property Prediction ("reaction yield") | Nguyen et al. (2020) ACS Catal 10(2):921-932 | RMSE (the lower the better) | - | - | 2.683 ('topk') | 2.652 ('topk') | - | - | - | - | | - | - | - | - | 1.936 ('topk') | 2.114 ('topk') | 3.247 ('topk') | 4.173 ('topk') | - | - | - | - | - |
| Fine-tuning Large Language Models for Chemical Text Mining | Action Sequence extraction | Specified in Suppl Table but not accessible at time of writing | BLEU | 49.5 ('30-shots') | 84.8 (fine-tuned with 1060 samples) | 65.0 ('60-shots') | - | - | - | - | 81.6(fine-tuned with 1060 samples) | 74.4(fine-tuned with 1060 samples) | 84.1(fine-tuned with 1060 samples) | - | - | - | - | - | - | - | - | - | - | - | - |

Values are extracted from corresponding reference listed in the first column. In red are numbers that are no better than at least one kind of GPT. In paratheses are prompting strategies or additional settings detailed in the corresponding litetature.